\documentclass[twocolumn,showpacs,pre,preprintnumbers,superscriptaddress,amsmath,amssymb]{revtex4}
\usepackage{graphicx}% Include figure files
\usepackage{dcolumn}% Align table columns on decimal point
\usepackage{bm}% bold math
\usepackage{amsmath,amssymb}
\newcommand{\be}{\begin{eqnarray}}
\newcommand{\ee}{\end{eqnarray}}

\begin{document}
\title{Spectral statistics, finite-size scaling and multifractal analysis of quasiperiodic chain with p-wave pairing}
\author{Yucheng Wang}
\affiliation{Beijing National
Laboratory for Condensed Matter Physics, Institute of Physics,
Chinese Academy of Sciences, Beijing 100190, China}
\author{Yancheng Wang}
\affiliation{Beijing National
Laboratory for Condensed Matter Physics, Institute of Physics,
Chinese Academy of Sciences, Beijing 100190, China}
\author{Shu Chen}
\thanks{schen@iphy.ac.cn}
\affiliation{Beijing National
Laboratory for Condensed Matter Physics, Institute of Physics,
Chinese Academy of Sciences, Beijing 100190, China}
\affiliation{Collaborative Innovation Center of Quantum Matter, Beijing, China}
\begin{abstract}
We study the spectral and wavefunction properties of a one-dimensional incommensurate system with p-wave pairing and unveil that the system demonstrates a series of  particular properties in its ciritical region. By studying the spectral statistics, we show that the bandwidth distribution and level spacing distribution in the critical region follow inverse power laws, which however break down in the extended and localized regions. By performing a finite-size scaling analysis, we can obtain some critical exponents of the system and find these exponents fulfilling a hyperscaling law in the whole critical region. We also carry out a multifractal analysis on system's wavefuntions by using a box-counting method and unveil the wavefuntions displaying different behaviors in the critical, extended and localized regions.

\end{abstract}
\maketitle
%%%%%%%%%%%%%%%%%%%%%%%%%%%%
\section{Introduction}
%%%%%%%%%%%%%%%%%%%%%%%%%%%%%
The localization to delocalization transition in one-dimensional (1D) quasiperiodic systems has attracted renewed attentions in recent years due to the experimental
progress in the physical realization of the quasiperiodic systems in ultracold
atomic systems and photonic crystals \cite{Roati,Deissler}. The 1D quasiperiodic systems have been studied in solid state physics to model the 1D electronic systems with incommensurate modulations of structures \cite{AA} and the Bloch electron in an incommenurate magnetic field \cite{Harper}. Different from the Anderson model with random disorders, the quasi-periodic systems can exhibit the loacalization to delocalization transition even in one dimensions,  which is well described by the Aubry-Andr\'{e} (AA) model \cite{AA}. As a prototype model of 1D quasiperiodic systems with the localization transition point exactly determined by a self-duality mapping, the AA model and its various extensions have been been extensively studied in past decades \cite{Wilkinson,Kivelson}.  By considering the interaction effect, recently it has also been applied to study the phenomenon of quantum many-body localization \cite{MBL1,MBL2}.

Motivated by recent studies of interplay of the disorder and topological superconductor (TSC) \cite{huse,pw,INS,Sarma,pw2}, the effect of 1D quasiperiodic (incommensurate) potentials on the topological superconductors has been explored \cite{M1,M2,Chen,WDS,hui}. Particularly, a 1D p-wave superconductor model with quasiperiodic potential was used to study the transition from TSC phase to Anderson localization induced by the incommensurate potential \cite{Chen,WDS}, and the localization transition point is analytically determined \cite{Chen}. In a recent work \cite{hui}, it was found that this model can be also used to describe a topologically non-trivial tight-binding model on the square lattice subjected to a non-Abelian gauge field. Through a number of numerical investigations, it was indicated that there exists a critical area before the system enters the localized phase. The critical phase is distinguished from the usual extended phase by different scaling behaviors of their wavefuntions \cite{hui} although both of them are topologically nontrivial superconductor states \cite{Chen}. Despite the fact that the localization transition point has been analytically determined by the close of excitation gap \cite{Chen}, it is still not clear whether the phase boundary between the critical phase and the usual extended phase can be determined from the spectrum properties? To clarify this point, we shall study the spectral statistics of the system and scrutinize whether the bandwidth distribution and level spacing distribution display different behaviors in the critical, extended and localized regions? On the other hand, although it was shown that the wavefuntions in the critical region display different scaling behaviors from that in the extended region, some important issues, including the critical exponents in the critical region and whether there exist some universal relations among these exponents, are not explored yet. Aiming to give answers to these questions, we shall carry out finite-size scaling analysis on the wavefuntions, which enables us to determine critical exponents of the system and verify the existence of a hyperscaling law among these exponents in the whole critical region.

The paper is organized as follows. In section II, we introduce the model and briefly review the method on solving the spectrum and wavefunctions. In section III, we study the spectral statistics of our model and find that the bandwidth distribution and the level-spacing distribution display an inverse power law (IPL) in the critical region. In section IV, we carry out a finite-size scaling analysis on system's wavefuntions to determine critical exponents of the system in the critical phase and unveil the existence of a hyperscaling law among these exponents. In section V, we perform a multifractal analysis on the wavefuntions by using a box-counting method. A brief summary is given in section VI.

%%%%%%%%%%%%%%%%%%%%%%%%
\section{Model and method}
%%%%%%%%%%%%%%%%%%%%%%%%%
We consider the model of a 1D p-wave superconductor in an incommensurate lattice described by
\begin{equation}
\begin{aligned}
&H = \sum_{i=1}^{L} [(-t\hat{c}^\dagger_{i}\hat{c}_{i+1} +\Delta\hat{c}_{i}\hat{c}_{i+1} +H.c.)+V_i \hat{n}_i]
\label{ham-1}
\end{aligned}
\end{equation}
with
\begin{equation}
 V_i=V \cos(2\pi\alpha i),
\label{tb1 }
\end{equation}
where $\hat{c}_{i}$ is a fermionic annihilation operator, $\hat{n}_i=\hat{c}^\dagger_{i}\hat{c}_{i}$ is the particle number operator, $V$ is the strength of the incommensurate potential, and $\alpha$ is an irrational number (we take $\alpha=\frac{\sqrt{5}-1}{2}$ in this work). For convenience, we shall take the hopping amplitude $t=1$ to be the unit of energy and set $0 <\Delta <1$ in the following calculation, which means $(t-\Delta) >0$. We will discuss the general situation in the appendix.
%, and $\Delta=0.5$.
When $\Delta=0$, the model (\ref{ham-1}) reduces to the well-known AA model \cite{AA}, for which a delocalization to localization transition occurs at the self-duality point $V=2t$ as long as $\alpha$ is an irrational number. In the presence of p-wave pairings ($\Delta \neq 0$), the model (\ref{ham-1}) exhibits a transition from a topological superconducting phase to a localized phase with the increase of $V$. The transition point at $V=2(t+\Delta)$ can be determined by the close of the excitation gap \cite{Chen,WDS}.   The system is in a localized phase when $V>2(t+\Delta)$, whereas the system is in an extended phase when $V<2(t+\Delta)$. The localized and extended phase can be distinguished by mean inverse participation ratio (MIPR) of wavefunctions of system's eigenstates \cite{Chen}. The MIPR tends to a finite number for the localized phase, whereas it tends to zero in the large $L$ limit for the extended phase.   In a recent work \cite{hui}, it was indicated that there exists a critical region at $2|t-\Delta|\leq V\leq 2(t+\Delta)$. In the critical region, the MIPR scales like $L^{-\eta}$ with $0<\eta<1$ \cite{hui}.

The Hamiltonian (\ref{ham-1}) can be diagonlized by using the Bogoliubove-de Gennes (BDG) transformation \cite{gennes,lieb,Lang}:
\begin{equation}
 \eta^\dagger_{n}=\sum_{i=1}^{L}[u_{n,i}\hat{c}^\dagger_{i}+v_{n,i}\hat{c}_{i}],
\label{tb2 }
\end{equation}
where $L$ is the number of lattice sites, and the diagonalized Hamiltonian is written as $H=\sum_{n=1}^{L}E_n(\eta^\dagger_{n}\eta_n-\frac{1}{2})$ with $E_n$ being the spectrum of the single quasi-particles. For the $n$th quasi-particle state $|\Psi\rangle=\eta^\dagger_n|0\rangle$, the equation $H|\Psi\rangle=E_n|\Psi\rangle$ gives the following explicit form
\begin{equation}
 -t(u_{n,i+1}+u_{n,i-1})+V_iu_{n,i}+\Delta(v_{n,i-1}-v_{n,i+1})=E_nu_{n,i},
\label{tb3}
\end{equation}
\begin{equation}
 t(v_{n,i+1}+v_{n,i-1})-V_iv_{n,i}+\Delta(u_{n,i+1}-u_{n,i-1})=E_nv_{n,i}.
\label{tb4}
\end{equation}
%Eq.(\ref{tb3}) and Eq.(\ref{tb4}) are independent of $n$.

For the irrational number $\alpha=\frac{\sqrt{5}-1}{2}$, which is also known as the inverse of the golden mean, it can be approached by the Fibonacci numbers via the relation
\be
 \lim_{\ell \rightarrow \infty}\frac{F_{\ell-1}}{F_{\ell}}=\alpha,
\ee
where the Fibonacci numbers $F_{\ell}$ are defined recursively by  $F_{\ell+1}=F_{\ell-1}+F_{\ell}$, with $F_0=F_1=1$ \cite{kohmoto,Kohmoto1983,Tang}.
%When $\alpha=F_{\ell-1}/{F_{\ell}}= K/L$, Eq. (\ref{ham-1}) is periodic with period $L$. We have $\psi_{n+L}=e^{ikL}\psi_n \; (n = 1,2, \cdots, L)$.
Numerically, we may successively change the system size $L=F_{\ell}$ to approach the irrational number. If we introduce a vector $|\Psi\rangle=
[u_1,v_1,u_2,v_2,\cdots,
u_L,v_L]^{T}$, solving Eq.(\ref{tb3}) and Eq.(\ref{tb4}) reduces to an eigenvalue problem of a
$2L\times 2L$ matrix \cite{kohmoto}:
\begin{equation}
 H =
\begin{pmatrix}
 A_1&B&0&\cdots& & &0 &C\\
B^{\dagger}&A_2&B&0&\cdots& & &0\\
0&B^\dagger&A_3&B&0&\cdots& &0\\
0&0&B^\dagger&A_4&B&0&\cdots&0\\
\vdots&\ddots&\ddots&\ddots&\ddots&\ddots&\ddots&\vdots\\
0& &\cdots &0 &B^\dagger &A_{L-2} &B &0\\
0& & &\cdots &0 &B^\dagger &A_{L-1} &B\\
C^\dagger& & & &\cdots &0 &B^\dagger &A_L
\end{pmatrix},
\label{num-1}
\end{equation}
where
\begin{equation}
 A_n= V\cos(2\pi\alpha i)
\begin{pmatrix}
 1 & 0\\
 0 & -1
\end{pmatrix},
\end{equation}
\be
 B=
\begin{pmatrix}
 -t & -\Delta\\
\Delta & t
\end{pmatrix},
\ee
and
\be
C=
\begin{pmatrix}
 -t & \Delta\\
-\Delta & t
\end{pmatrix}
\ee
for the system with periodic boundary conditions (PBC),
or
\be
C=
\begin{pmatrix}
 0 & 0\\
0 & 0
\end{pmatrix}.
\ee
for the system with open boundary conditions (OBC).

%%%we use $L=6765$, which corresponds to $F_{19}$.
%%%%%%%%%%%%%%%%%%%%%%%%%%%%%%%%%%%%%%%%%%%%%%
\begin{figure}
\includegraphics[height=120mm,width=80mm]{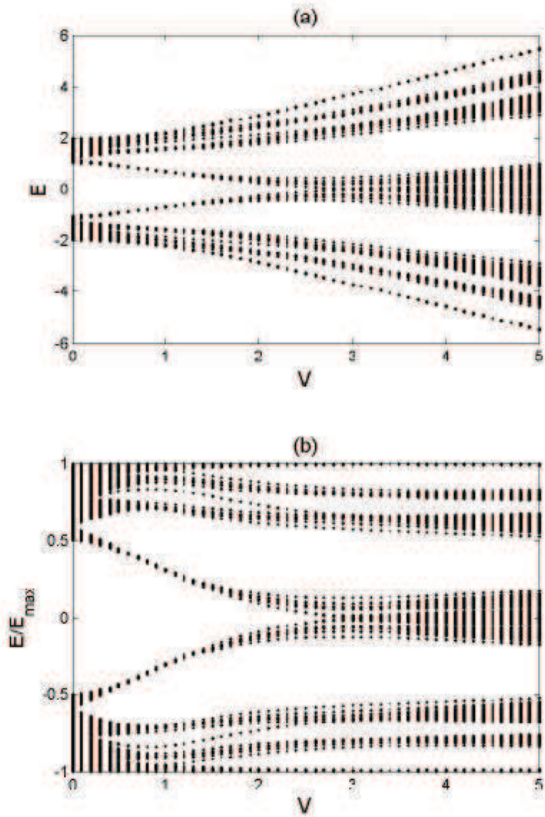}
\caption{\label{000}
  (a) Energy as a function of potential strength $V$. (b) Energy spectrum normalized by the maximized energy $E_{max}$ for various potential strength $V$.  The parameters are $\Delta=0.5$, $t=1$ and $L=1597$.}
\end{figure}
%%%%%%%%%%%%%%%%%%%%%%%%%%%%%%%%%%%%%%%%%%%%%%

%%%%%%%%%%%%%%%%%%%%%%%%%%%%
\section{Level distribution and band width distribution}
%%%%%%%%%%%%%%%%%%%%%%%%%%%%%
 The statistical description of the energy levels is an important tool in the study of quantum systems with complex structures, including nuclear systems, quantum chaos, and condensed matter systems \cite{PhyRep}. It is known that the distribution of energy levels of a disordered quantum system exhibit different properties in extended and localized phases. For the the special case of our model (\ref{ham-1}) with $\Delta=0$, the level statistics of the AA model has been well studied \cite{Pet,Fujita}. It is found that there exist distinctive level distributions, i.e., the Poisson, IPL, and cosine-band-like behaviors, depending on whether the system is localized, critical and extended \cite{Fujita,Pet}. While the critical phase only occurs at the self-duality point $V=2t$ for the AA model, there exists a critical region for our model in the presence of the pairing term.
To unveil properties of spectral statistics in the critical region, we shall investigate level-spacing distribution \cite{Pet,Fujita,PRE} of our model.

To get an intuitive picture of the spectral distribution, firstly we display the spectrum of the model (\ref{ham-1}) with $\Delta=0.5$ and $t=1$ under PBC in Fig.~\ref{000}(a) for various potential strength $V$. We also display the scaled spectrum of the system in Fig.~\ref{000}(b), in which the spectrum is normalized by the maximized energy $E_{max}$ of the system with a given $V$. It is clear that the spectrum is symmetric about the zero energy due to the existence of particle-hole symmetry, and the ground state of the system corresponds to the state with all the negative energy levels filled. One can see that the excitation gap closes at the localization transition point $V=3$ \cite{Chen}, however no obvious change is found in the spectrum around the other transition point $V=1$. Although the boundary between the critical region and extended region is not discernable, it seems that the spectrum in the critical region displays a self-similar structure.

%%%%%%%%%%%%%%%%%%%%%%%%%%%%%%%%%%%%%%%%%%%%%%
\begin{figure}
\includegraphics[height=60mm,width=75mm]{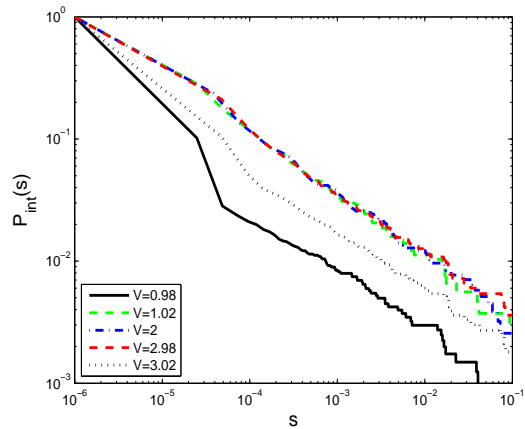}
\caption{\label{002}
  Integrated level-spacing distribution of system with different incommensurate potential $V$ ($V=0.98, 1.02, 2, 2.98, 3.02$) for the rational approximation $\alpha=\frac{10946}{17711}$ of the golden mean. Other parameters are $\Delta=0.5$ and $t=1$.}
\end{figure}

To get a more quantitative description, next we study the level-spacing distribution of the model. For the AA model, the level-spacing distribution is found to fulfill the IPL at the critical point $V=2t$ \cite{Fujita,Pet}. To check whether the the level-spacing distribution of our model fulfills the IPL in the whole critical region, it is convenient to calculate the integrated level-spacing distribution (ILSD) \cite{Pet}
\begin{equation}
 p_{int}(s)=\int_s^{\infty}p(s')ds',
 \label{tb5}
\end{equation}
whose derivative $p(s)=-dp_{int}/ds$ gives the distribution density of the nearest-neighbor level spacing $s$, and $p_{int}(s)$ determines the probability density of the energy gaps larger than the spacing size $s$. We plot the ILSD for the system with $L=17711$ (corresponding to $F_{19}$) for $V=0.98, 1.02, 2, 2.98, 3.02$ under PBC in the Fig.~\ref{002}, in which we have introduced a cutoff $s_{min}=10^{-6}>0$ and normalized the function $p_{int}$. In the critical region, the ILSD clearly displays an IPL
\begin{equation}
 p_{int}(s)\sim s^{1-\rho},
 \label{tb6}
\end{equation}
and thus the level-spacing distribution behaves as
\begin{equation}
 p(s)\sim s^{-\rho},
 \label{tb7}
\end{equation}
where $\rho=1.5177 \pm 0.0029$ , $1.5199 \pm 0.0048$, and $1.4861 \pm 0.0024$ for $V=1.02$£¬$2$ and $2.98$, respectively.
As the level-spacing distribution in the critical region fulfills the IPL, it obviously stays away from the IPL at the extended ($V=0.98$) and localized phase ($V=3.02$) as shown in Fig.~\ref{002}. The distribution of IPL has been taken as a signature reflecting the self-similarity of the structure of spectrum \cite{Fujita,Pet}.
%%%%%%%%%%%%%%%%%%%%%%%%%%%%%%%%%%%%%%%%%%%%%
\begin{figure}
\includegraphics[height=180mm,width=90mm]{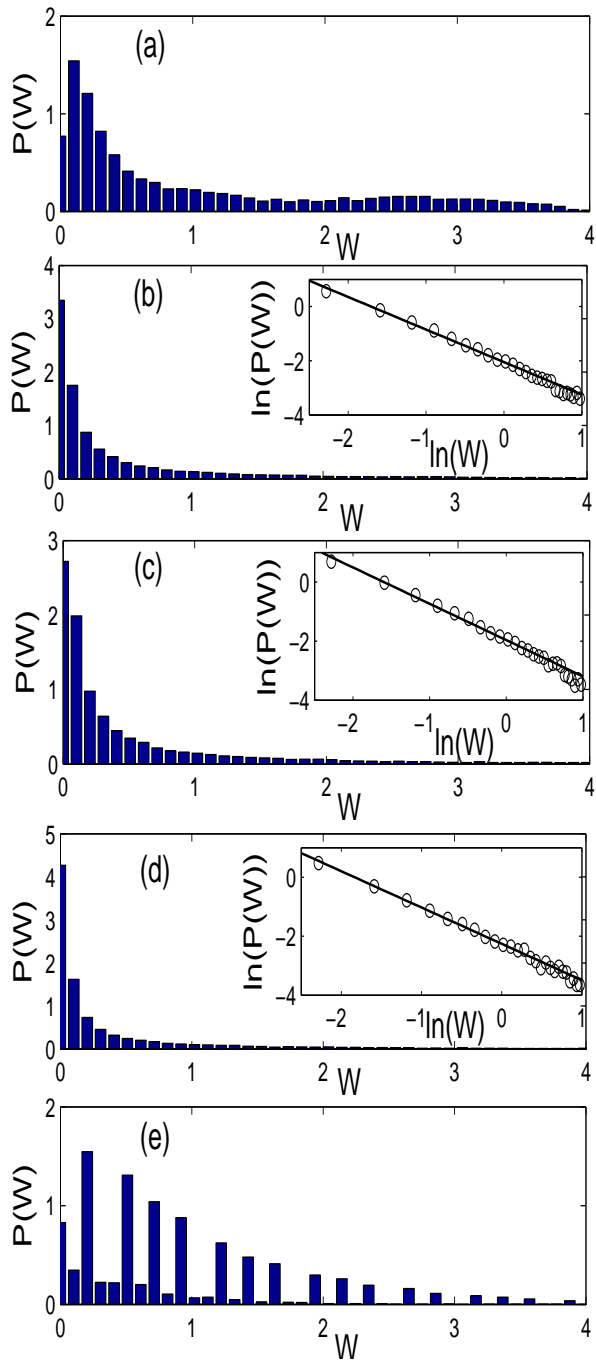}
\caption{\label{001}
  The bandwidth distribution $P(w)$ for (a) $V=0.98$ (b) $V=1.02$, (c) $V=2$, (d) $V=2.98$ and (e) $V=3.02$ by fixing $L=17711$. The legend of (b), (c) and (d) refer to the logarithm of the corresponding bandwidth distribution.}
\end{figure}

 Next we consider the bandwidth distribution $P(W)$, where $W$ is the normalized bandwidth, which is the bandwidth divided by the average bandwidth of all $2L$ bands.  The bandwidth distribution of the AA model has been studied in Ref.\cite{Kohmoto1983,Tang,PRE}. It is found that the bandwidth distribution of the AA model at the critical point displays quite different behavior from that in the extended and localized phases \cite{Kohmoto1983,Tang,PRE}.
 To calculate the bandwidth, one may consider the system described by Eq. (\ref{ham-1}) with $\alpha=F_{\ell-1}/F_{\ell}$ being periodic with a period $L=F_{\ell}$, which can be solved by using the Bloch condition, e.g., $u_{j+L}=e^{ikL}u_j \; (j = 1,2, \cdots, L)$, where $k$ is the Bloch index. For a fixed value of $k$, the system can be solved by diagonalizing the martix (\ref{num-1}) with
\be
C=
\begin{pmatrix}
 -t & \Delta\\
-\Delta & t
\end{pmatrix}
e^{-ikL} .
\ee
Its eigenvalues form $2F_{\ell}$ energy bands as $k$ is varied in the first Brillouin zone $[-\pi/F_{\ell}, \pi/F_{\ell}]$.
In Fig.~\ref{001}, we display the bandwidth distribution $P(W)$ for $V=0.98$, $1.02$, $2$, $2.98$ and $3.02$ by fixing $\Delta=0.5$, $t=1$ and $L=17711$ . From Fig.~\ref{001}(b)-(d), we see that $P(W)$ follows an IPL
\begin{equation}
P(W) \sim W^{\kappa},
\end{equation}
where $\kappa = -1.201 \pm 0.018$ for $V=1.02$,  $\kappa=-1.229 \pm 0.024$ for $V=2$, and $\kappa=-1.231 \pm 0.025$ for $V=2.98$ in the critical region. However, $P(W)$ follows different laws in the extended region and the localized region as shown in Fig.~\ref{001}(a) and (e).

%%%%%%%%%%%%%%%%%%%%%%%%%%%%%
%%%%%%%%%%%%%%%%%%%%%%%%%%%%%%%%%%%%%%%%%%%%%%
\begin{figure}[t]
 \includegraphics[height=100mm,width=80mm]{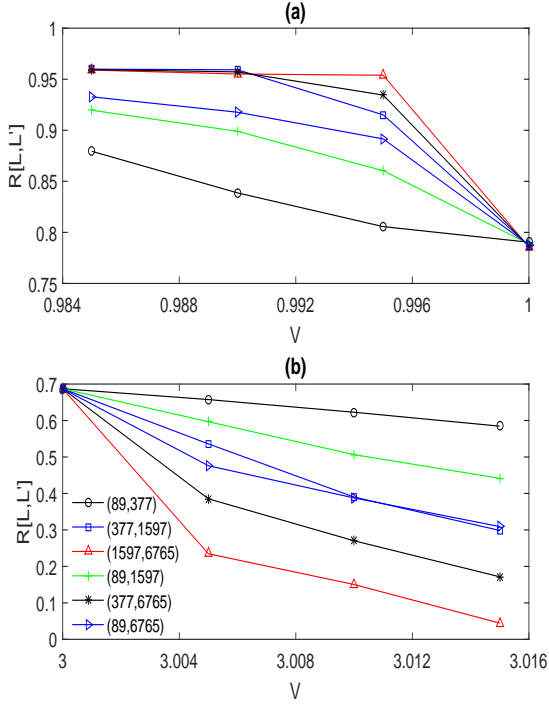}
\caption{\label{gamma}
The plots of $R[L,L^{'}]$ versus $V$ for several pairs of ($L, L^{'}$) by fixing $t=1$, $\Delta=0.5$ and $q=2$. (a) corresponds to the left boundary of the critical area, describing the transition from the extended phase to critical phase. (b) corresponds to the right boundary of the critical area, describing the transition from the critical phase to localized phase.}
\end{figure}
%%%%%%%%%%%%%%%%%%%%%%

%%%%%%%%%%%%%%%%%%%%%%%%%%%%
\section{A finite-size scaling analysis}
\label{scaling}
%%%%%%%%%%%%%%%%%%%%%%%%%%%%%%%

Given that $|\Psi_n \rangle=[u_{n,1},v_{n,1},\cdots,u_{n,L},v_{n,L}]^T$ is a normalized eigenvector corresponding to the $n$th eigenvalue, where $L$ is the system size, we can define the mean 'sum over states' as \cite{YKY}
\begin{equation}
 Z_L(q)=\frac{1}{L}\sum_{n=1}^{L}\sum_{i=1}^{L}(|u_{n,i}|^{2q}+|v_{n,i}|^{2q}),
\label{ZL}
\end{equation}
where $q$ is a real parameter. The generalized participation ratio is defined by using $Z_L(q)$ as
\begin{equation}
 I_L(q)={Z_L(q)}^{1/(1-q)}.
\label{lL}
\end{equation}
When $q=2$, $I_L(2)$ is just the conventional participation ratio, which has been widely used in the study of the Anderson localization in disordered systems \cite{Thouless} and incommensurate systems \cite{PRA}. For the AA model, it has been indicated that $I_L(q)/L$ has similar properties to $\langle M^2 \rangle$ in the Ising model, where $M$ is the instantaneous magnetization of the system composed of $L$ spins and $\langle M^2 \rangle$ is the time average of $M^2$ \cite{YKY}.  Then we can define the quantity:
\begin{equation}
 \sigma_L(q)=(I_L(q)/L)^{1/2}.
\label{sigma0}
\end{equation}
Denote $I(q)= I_{\infty}(q)$ and $\sigma(q)=\sigma_{\infty}(q)$ as the thermodynamic limits of $I_L(q)$ and $\sigma_L(q)$  ($L \rightarrow\infty$), respectively. %$\sigma$ corresponding to the average magnetization.
Near a phase transition point $V_c$,  we can define three critical exponents as \cite{YKY}:
\begin{equation}
 \xi \sim |\delta V|^{-\nu}, \label{xi}
\end{equation}
\begin{equation}
 I(q) \sim (\delta V)^{-\gamma}, \label{lq}
\end{equation}
\begin{equation}
 \sigma(q) \sim (-\delta V)^{\beta},
 \label{sigm}
%\label{expontent}
\end{equation}
where $\xi$ is the correlation length or the localization length, and $\delta V=(V-V_c)/V_c$, where $V \leq V_{c_1}=2(t-\Delta)$ when we consider the extended-critical transition and $V \geq V_{c_2}=2(t+\Delta)$ when we consider the critical-localized transition. Near the critical point, we assume the following finite-size scaling relationship for a finite system:
\begin{equation}
 \sigma_L(q)^2L^{1-\gamma/\nu}=F(L^{1/\nu}(\delta V)),
\label{function}
\end{equation}
where F(x) is the scaling function. For convenience, we set $q=2$ in the following calculation.

At the critical point $V=V_c$, we have $\delta V=0$, and thus Eq.(\ref{function}) reduces to $\sigma_L^2=F(0)L^{\gamma/\nu-1}$. Then we can define a function of two size-variables by
\begin{equation}
 R[L,L^{'}]=\frac{log(\sigma^2_L/\sigma^2_{L^{'}})}{log(L/L^{'})}+1.
\label{function2}
\end{equation}
It is clear that the above function equals to $\gamma/\nu$ at $V=V_c$ for any pair ($L,L^{'}$) provided that $L$ and $L^{'}$ are large enough. In Fig.~\ref{gamma}, we display the change of $R[L,L^{'}]$ as a function of $V$ for different pairs of $L$ and $L^{'}$ \cite{YKY}, where the PBC is used. We can determine $\gamma/\nu$ and $V_c$ from the crossing point in the Fig.~\ref{gamma}. For the system with $\Delta=0.5$ and $t=1$, we can get the critical strength $V_{c_1}=1.00$ and the corresponding critical exponent $\gamma/\nu=0.7867$ from Fig.~\ref{gamma}(a), which is consistent with $V_{c_1}=2(t-\Delta)$. Similarly, we get the critical strength $V_{c_2}=3.00$ and the corresponding critical exponent $\gamma/\nu=0.6871$ from Fig.~\ref{gamma}(b), which is consistent with $V_{c_2}=2(t+\Delta)$. By this way, we can numerically determine both the phase boundaries of the critical region very precisely.
%%%%%%%%%%%%%%%%%%%%%%%%%%%%%%%%%%%%%%%%
\begin{figure}[t]
 \includegraphics[height=110mm,width=80mm]{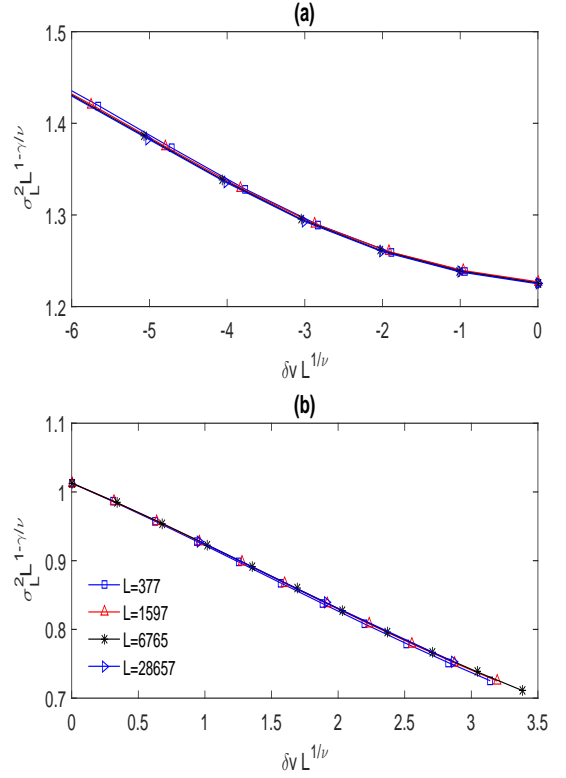}
\caption{\label{nu} $\sigma_L(q)^2L^{1-\gamma/\nu}$ versus $L^{1/\nu}(\delta V)$ for different values of $L$ with (a) $V\leq V_c=2|t-\Delta|$ and (b) $V\geq V_c=2(t+\Delta)$. Different lines are superposed together when we set $\nu=1$. Here we have taken $t=1$, $\Delta=0.5$ and $q=2$.}
\end{figure}
%%%%%%%%%%%%%%%%%%%%%%

The exponent $\nu$ can be determined by plotting of $\sigma_L(q)^2L^{1-\gamma/\nu}$ versus $L^{1/\nu}(\delta V)$ for different values of $L$ with $V\leq 2|t-\Delta|$ in Fig.~\ref{nu}(a) and $V\geq 2(t+\Delta)$ in Fig.~\ref{nu}(b), respectively.  Close to the transition points, it is shown that lines corresponding to different L superpose together if we set the exponent $\nu=1$, which indicates $\nu=1$ at both the left boundary and right boundary of the critical area.

By using Eq. (\ref{xi}), we can rewrite Eq. (\ref{lq}) and Eq. (\ref{sigm}) in terms of $\xi$ as
\begin{equation}
 I(q) \sim \xi^{\frac{\gamma}{\nu}}, \label{lq2}
\end{equation}
\begin{equation}
 \sigma(q) \sim \xi^{-\frac{\beta}{\nu}}. \label{sigm2}
%\label{expontent}
\end{equation}
At the critical point, $\xi \sim L$,  so we have \cite{Isakov}
\begin{equation}
 I(q) \sim L^{\frac{\gamma}{\nu}}, \label{lq3}
\end{equation}
\begin{equation}
 \sigma(q) \sim L^{-\frac{\beta}{\nu}}. \label{sigm3}
%\label{expontent}
\end{equation}
As long as Eq. (\ref{lq3}) and Eq. (\ref{sigm3}) hold true, we can directly get the following hyperscaling law from Eq. (\ref{sigma0}):
\begin{equation}
 \frac{2\beta}{\nu}+\frac{\gamma}{\nu}=1. \label{law}
\end{equation}
To verify this hyperscaling law numerically, we plot $ln(I_{L}(2))$ and $ln(\sigma_{L}(2))$ versus $ln(L)$ in Fig.~\ref{law}.
As shown in  Fig.~\ref{law} (a) and (b), we observe $ln(I_{L}(2))$ and $ln(\sigma_{L}(2))$ as a linear function of $ln(L)$ for $V=1$, $1.5$, $2.5$ and $3$, indicating that Eq. (\ref{lq3}) and Eq. (\ref{sigm3}) are fulfilled in the whole critical region $1 \leq V \leq 3$. From the slope of the straight line in  Fig.~\ref{law} (a), we can obtain  ${\gamma}/{\nu}=0.7871$, $0.7390$, $0.7381$, and $0.6871$ corresponding to $V=1$, $1.5$, $2.5$ and $3$, respectively.
Similarly, we can get ${\beta}/{\nu}=0.1065$, $0.1305$, $0.1309$ and $0.1565$ corresponding to $V=1$, $1.5$, $2.5$ and $3$, respectively, from the slope of the straight line in  Fig.~\ref{law} (b). It is clear that the hyperscaling law is fulfilled for all these states in the critical region. As a comparison, we also give data for the system out of the critical region in Fig.~\ref{law} (a) and (b). For $V=0.99$ and $V=3.01$, we can see that $ln(I_{L}(2))$ and $ln(\sigma_{L}(2))$ are no longer linear functions of $ln(L)$. Consequently, Eq. (\ref{lq3}) and Eq. (\ref{sigm3}) are not fulfilled in the extended and localized region.
%and no a hyperscaling law exists.
%%%%%%%%%%%%%%%%%%%%%%%%%%%%%%%%%%%%%%%%
\begin{figure}[t]
 \includegraphics[height=110mm,width=90mm]{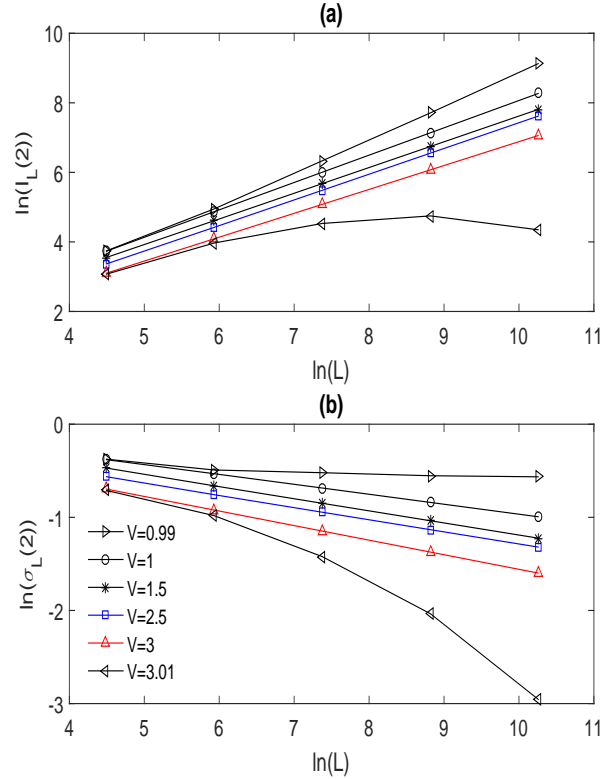}
\caption{\label{law}
(a) $ln(I_L(2))$ as a function of $ln(L)$ for $V = 0.99$, $1$, $1.5$, $2.5$, $3$ and $3.01$, respectively. (b) $ln(\sigma_L(2))$ versus $ln(L)$ for $V=0.99$, $1$, $1.5$, $2.5$, $3$ and $3.01$, respectively. We can obtain  $\frac{\gamma}{\nu}$ and $\frac{\beta}{\nu}$ from (a) and (b) for different $V$ in the critical region and the relationship between $ln(I_L(2))$ or $ln(\sigma_L(2))$ and $ln(L)$ isn't linear in the extended and localized region. Here we fix $t=1$ and $\Delta=0.5$.}
\end{figure}
%%%%%%%%%%%%%%%%%%%%%%

%%%%%%%%%%%%%%%%%%%%%%%%%%%%
\section{multifractal analysis of wave functions}
%%%%%%%%%%%%%%%%%%%%%%%%%%%%%%%
Next we perform a multifractal analysis of the wave functions of the model (\ref{ham-1}) by using the box-counting method \cite{AL,Martin}, which was applied to study the AA model \cite{AL}. For the AA model, it was found that the wavefuntions exhibit multifractal fluctuations extending to all length scales at the critical point ($V_c=2$). When $V>V_c$, the localized states manifest multifractal feature up to the localization length. On the other side with $V<V_c$, the extended states manifest multifractal features up to the correlation length. For our model, it would be interesting to study whether the multifractal features exist in the whole critical region and how these multifractal properties change with different lattice sizes.
%and how to define the localization length and correlation length when the system deviates from the critical regime.

In the above section, we have determined the critical exponent $\nu=1$ at both the transition point $V_{c_1} =2(t - \Delta)$ and $V_{c_2} = 2(t + \Delta)$. Therefore, from Eq.(\ref{xi}), we can get the correlation length $\xi_c = a \frac{V_{c_1}}
{V_{c_1}-V}$, when $V$ approaches $V_{c_1} = 2(t-\Delta)$ from $V<V_{c_1}$, and the localization length $l_c=b \frac{V_{c_2}}{V-V_{c_2}}$, when $V$ approaches $V_{c_2}=2(t+\Delta)$ from $V>V_{c_2}$, where $a$ and $b$ are undetermined constants. When $\Delta \rightarrow 0$, this system becomes the AA model, and we can obtain $a=b=1$ by comparing $\xi_c$ and $l_c$ with the localization length and correlation length of the AA model \cite{AA,AD,Suslov}, given by $l_c=\frac{2}{V-2}$ and $\xi_c=\frac{2}{2-V}$, respectively.

%%%%%%%%%%%%%%%%%%%%%%%%%%%%
%\subsection{multifractal analysis of wave functions}
%%%%%%%%%%%%%%%%%%%%%%%%%%%%%%%
Then we consider multifractal properties of our model. Given a wave function defined over lattice size $L$ divided into $L/l$ segments of length $l$, we define a quantity
\begin{equation}
 \chi_j(q)=\sum_{n=1}^{L/l}[\sum_{i=(n-1)l+1}^{nl}(u_{j,i}^2+v_{j,i}^2)]^q,
\label{chi}
\end{equation}
and the average of them
\begin{equation}
 \chi(q)=\frac{1}{L}\sum_{j=1}^{L}\chi_j(q).
\label{chi2}
\end{equation}
as a function of $l$, where $j$ corresponds the $jth$ eigenstate.
%From the behavior of $\chi(q)$ for small $l$, we can get the value of the exponent $\tau(q)$ which provides a characterization of the eventual fractal singularities of the probability distribution \cite{AL}.
Multifractality is characterized by a power-law behavior of  $\chi(q)  \sim  (l/L)^{\tau (q)}$ with the exponent $\tau(q)$ determining the multifractal dimensions $D_q = \tau(q)/(q-1)$ \cite{Martin,boris,Martin14,Mirlin}.
We now consider the case of $q=2$ and discuss some detail properties of $\tau(2)$ as a function of the length-scale considered. Here $\tau(2)$ equals to the correlation dimension \cite{boris}, i.e., $\tau(2)=D_2$. While $D_2$ tends to $0$ for a localized  state and tends to $1$ for an extended state. The wave functions are multifractals if $0<D_2<1$.
%%%%%%%%%%%%%%%%%%%%%%%%%%%%%%%%%%%%%%%%%%%%%%
\begin{figure}
\includegraphics[height=120mm,width=80mm]{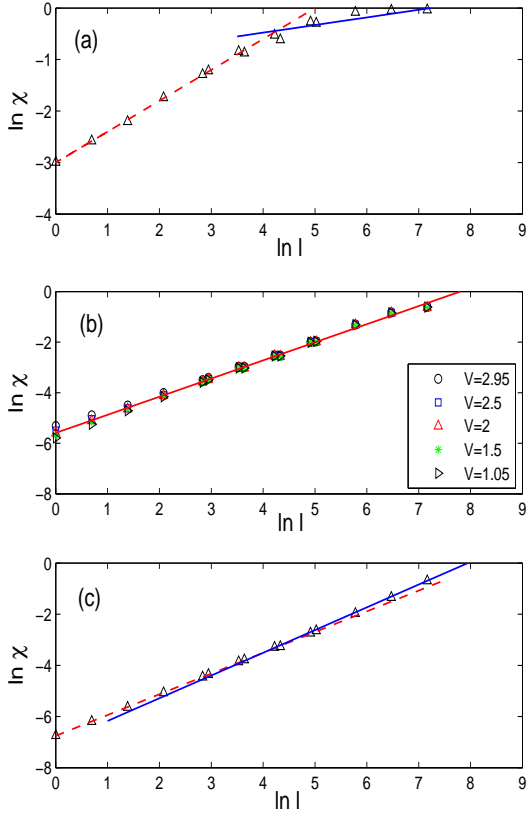}
\caption{\label{01}
  $ln\chi(q=2)$ as a function of $lnl$ for $L=2584$ (corresponds to $F_{17}$), where $l$ is the box size used in Eq.(\ref{chi}) (a) corresponds to a localized phase with $V=3.05$. (b) corresponds to the critical case, and the legend refers to different incommensurate potential strength $V=2.95, 2.5, 2, 1.5, 1.05$. Here we only plot the straight line about $V=2$. (c) corresponds to an extended phase with $V=0.95$.}
\end{figure}
%%%%%%%%%%%%%%%%%%%%%%%%%%%%%

In Fig.~\ref{01}, we display the change of  $ln\chi$ as a function of $lnl$ for different quasi-disorder strength $V$ by using PBC.
Fig.~\ref{01}(a) refers to $V=3.05$ corresponding to localized states. When the length $l$  is smaller than the localization length $l_c$, we see that $\chi(2)$ follows the power law $\chi(2) \sim l^{D_2}$, where $D_2=0.602\pm 0.014$ can be determined from the slope of the dashed line in Fig.~\ref{01} (a). On the other hand, the data for lengths larger than $l_c$ can be approximated by a line with the slope $D_2\approx 0$. Here a crossover is clearly present by changing the length $l$.
In Fig.~\ref{01}(b) we show the critical case with $V=1.05, 1.5, 2, 2.5$, and $2.95$.  No crossover is found for these cases and the wave functions are multifractals to all scales.  It is found that $ln\chi$ is a linear function of $lnl$ with the slope of the straight line given by $D_2=0.69$ to
$D_2=0.74$ corresponding to $V=1.05$ to $V=2.95$. In Fig.~\ref{01}(c) we show the extended case with $V=0.95$. In this case there is a crossover point.  Below this point, the slope is given by $D_2=0.808\pm 0.005$, exhibiting the multifractal feature, and above it $D_2\approx 1$, corresponding to an extended state without self-similarity.

%%%%%%%%%%%%%%%%%%%%%%%%%%%%%%%%%%%%%%%%%%%%%%
\begin{figure}
\includegraphics[height=110mm,width=80mm]{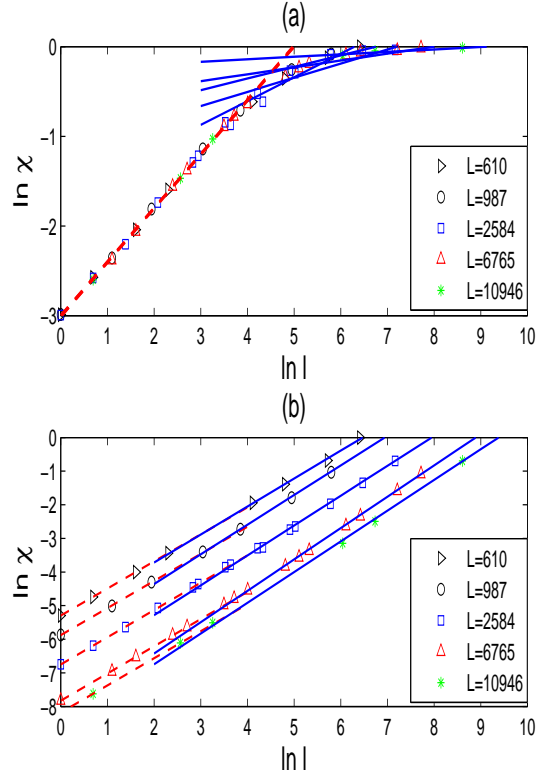}
\caption{\label{02}
  $ln\chi(q=2)$ as a function of $lnl$ for $L=610$ (corresponds to $F_{14}$), $L=987$ (corresponds to $F_{15}$), $L=2584$ (corresponds to $F_{17}$), $L=6765$ (corresponds to $F_{19}$), and $L=10946$ (corresponds to $F_{20}$), where $l$ is the box size used in Eq.(\ref{chi}). (a) corresponds to a localized phase with $V=3.05$. (b) corresponds to an extended phase with $V=0.95$. The legend refers to different lattice sizes.}
\end{figure}
%%%%%%%%%%%%%%%%%%%%%%%%%%%%%

We further study multifractal properties of our model with different lattice sizes.
In Fig.~\ref{02}(a), we display the change of  $ln\chi$ as a function of $lnl$ for the system with various $L$ and $V=3.05$ in the localized region. For $l<l_c$, one can observe $ln\chi$ being a linear function of $lnl$, which completely superposes together for different $L$  with an identical slope $D_2=0.602$ suggesting that the system exhibits multifractal feature. For $l>l_c$ with $l_c={3}/(3.05-3)$, the slopes of the straight lines are different for different lattice size $L$ and the slopes decrease to $0$ when enlarging $L$. In Fig.~\ref{02}(b), we display the change of  $ln\chi$ as a function of $lnl$ for the system with various $L$ and $V=0.95$ in the extended region. For $l<\xi_c$, we observe that $ln\chi$ versus $lnl$ are described by a series of parallel lines with the same slope $D_2=0.808$. When $l>\xi_c$, we find the slopes are different for different lattice size $L$ and the slopes increase to $1$ when enlarging $L$, where $\xi_c={1}/(1-0.95)$.

%%%%%%%%%%%%%%%%%%%%%%%%%%%
\section{Summary}
%%%%%%%%%%%%%%%%%%%%%%%%%%%%%%%
%\textit{Summary.-}
In summary, we have explored spectral statistics, a finite-size scaling and multifractal analysis of the 1D incommensurate system with p-wave pairing, which can be in the extended, critical, or localized phase depending on the strength of incommensurate potential. Our results indicate that the bandwidth distribution and level spacing distribution in the critical region fulfill inverse power laws, which no longer survive in the localized and extended region. By carrying out a finite-size scaling on the wavefuntions of the system, we can determine both the transition point from the localized phase to critical phase and that from the extended phase to critical phase. We have also acquired the critical exponents of this system and found the existence of a hyperscaling law among these exponents in the whole critical region. By using a box-counting method, we carry out multifractal analysis and unveil the wavefuntions displaying different behaviors in the critical, extended and localized regions. When the system is in the critical region, the wavefuntions are found to present multifractal behaviors at all length scales. On the other hand, if the system is in the localized (extended) phase near the phase transition point, the multifractal behaviors can only be observed below the length scale of the localization (correlation) length.
%with the counting-box size less than the localization (correlation) length.

%%%%%%%%%%%%%%%%%%%%%%%%%%%

%%%%%%%%%%%%%%%%%%%%%%%%%%%%%%%
%\textit{Acknowledgments.-}
%%%%%%%%%%%%%%%%%%
\begin{acknowledgments}
 We thank  Haiping Hu for helpful discussions. The work is supported by NSFC under Grants No. 11425419, No. 11374354 and No. 11174360, and the Strategic Priority Research Program (B) of the Chinese Academy of Sciences  (No. XDB07020000).
\end{acknowledgments}

\appendix
%%%%%%%%%%%%%%%%%%%%%%%%%%%%
\section{Symmetry analysis of  model}
%%%%%%%%%%%%%%%%%%%%%%%%%%%%%%%
Through a Jordan-Wigner transformation, the model given by Eq.(1) in the main text corresponds to a transverse XY model with an irrationally modulated transverse field \cite{Fisher,Young,Chen}:
\begin{equation}
\hat{H}=-\sum_i[J_x\sigma^x_i\sigma^x_{i+1}+J_y\sigma^y_i\sigma^y_{i+1}]+\sum_ih_i\sigma^z_i,
\end{equation}
where $J_x=(t+\Delta)/2$, $J_y=(t-\Delta)/2$, $h_i=-V_i/2$ and $\sigma^x_i=(\hat{c}^\dagger_{i}+\hat{c}_i)exp(-i\pi\sum_{j=1}^{i-1}\hat{c}^\dagger_{j}\hat{c}_j)$. We rotate the system on $z$ axis with $\pi/2$, which leads to
\begin{equation}
\begin{aligned}
&\hat{H}'=e^{i\frac{\pi}{2}\sum_iS^z_i}\hat{H}e^{-i\frac{\pi}{2}\sum_iS^z_i}\nonumber\\
&= e^{i\frac{\pi}{4}\sum_i\sigma^z_i}\hat{H}e^{-i\frac{\pi}{4}\sum_i\sigma^z_i}\nonumber\\
&= -\sum_i[J_x\sigma^y_i\sigma^y_{i+1}+J_y\sigma^x_i\sigma^x_{i+1}]+\sum_ih_i\sigma^z_i.
\end{aligned}
\end{equation}
It is clear that the above unitary transformation gives $U^{-1}H(\Delta)U=H(-\Delta)$ with $U=e^{-i\frac{\pi}{2}\sum_iS^z_i}$.

Introducing $\phi_{n,i}=(u_{n,i}+v_{n,i})$ and $\psi_{n,i}=(u_{n,i}-v_{n,i})$,  we can rewrite the BDG transformation (Eq.(3) in the main text) as
\begin{equation}
 \eta^\dagger_{n}=\frac{1}{2}\sum_{i=1}^{L}[(\phi_{n,i}+\psi_{n,i})\hat{c}^\dagger_{i}+(\phi_{n,i}-\psi_{n,i})\hat{c}_{i}].
\label{tb9 }
\end{equation}
In terms of ($\phi, \psi$), Eq.(4) and (5) in the main text can be represented as
\begin{equation}
 (\Delta-t)\psi_{n,i+1}+V_i \psi_{n,i}-(\Delta+t)\psi_{n,i-1}=E_n\phi_{n,i},
\label{tb10}
\end{equation}
\begin{equation}
 -(\Delta+t)\phi_{n,i+1}+V_i \phi_{n,i}+(\Delta-t)\phi_{n,i-1}=E_n\psi_{n,i}.
\label{tb11}
\end{equation}
By solving the above equations, we can diagonalize the Hamiltonian and obtain all its eigenvalues $E_n$ and eigenstates denoted by the vectors $|\Psi_n\rangle= [\psi_{n,1},\phi_{n,1},\psi_{n,2},\phi_{n,2},\cdots, \psi_{n,L},\phi_{n,L}]^{T}$. If we make a transformation $H(\Delta) \rightarrow H(-\Delta)$ and $\psi_{n,i}\rightarrow \phi_{n,i}$, $\phi_{n,i}\rightarrow \psi_{n,i}$, where $i=1,2,\cdots,L$, then Eq.(\ref{tb10}) becomes Eq.(\ref{tb11}) and Eq.(\ref{tb11}) becomes Eq.(\ref{tb10}). To be detailed, given that $H(\Delta)|\Psi_n\rangle=E_n|\Psi_n\rangle$,
%with $|\Psi_n\rangle=[\psi_{n,1},\phi_{n,1},\psi_{n,2},\phi_{n,2},\cdots, \psi_{n,L},\phi_{n,L}]^{T}$,
we have $U^{-1}H(\Delta)UU^{-1}|\Psi_n\rangle=U^{-1}E_n|\Psi_n\rangle$, which gives $H(-\Delta)|\Psi^{'}_n\rangle=E_n|\Psi^{'}_n\rangle$ with $|\Psi^{'}_n\rangle=U^{-1}|\Psi_n\rangle=[\phi_{n,1},\psi_{n,1},\phi_{n,2},\psi_{n,2},\cdots, \phi_{n,L},\psi_{n,L}]^{T}$. Now it is clear that the generalized partition ration defined by Eq.(18) in the main text and the MIPR defined by \cite{Chen,hui}
\begin{equation}
MIPR=\frac{1}{L}\sum_{n=1}^{L}\sum_{i=1}^{L}(|\phi_{n,i}|^{4}+|\psi_{n,i}|^{4}),
\end{equation}
are invariant under the unitary transformation. As the extended, critical and localized phase are characterized by the generalized partition ration or the MIPR, we can directly get properties of the system with $\Delta<0$ from the system with $\Delta>0$.

In this paper we numerically determine the critical region range of this system that $2(t-\Delta)\leq V \leq2(t+\Delta)$ with $t>0$, $\Delta>0$ and $(t-\Delta)>0$. From the above discussion, we can determine the critical region range $2(t-|\Delta|)\leq V \leq2(t+|\Delta|)$ when $t>0, (t-|\Delta|)>0$. If we make some replacements $t\rightarrow -t$, $\psi_{n,i}\rightarrow -\phi_{n,i}$, $\phi_{n,i}\rightarrow -\psi_{n,i}$ if $i$ is odd and $\psi_{n,i}\rightarrow \phi_{n,i}$, $\phi_{n,i}\rightarrow \psi_{n,i}$  if $i$ is even in Eq.(\ref{tb10}) and Eq.(\ref{tb11}), then Eq.(\ref{tb10}) becomes Eq.(\ref{tb11}) and Eq.(\ref{tb11}) becomes Eq.(\ref{tb10}). Consequently, when we change $t$ to $-t$, the extended and localized properties of this system aren't changed. Therefore we can determine the critical region $2(|t|-|\Delta|)\leq V \leq2(|t|+|\Delta|)$ with $(|t|-|\Delta|)>0$ just from the parameter region with $t>0$ and $\Delta>0$. In Ref.\cite{hui}, the authors make a local replacement about Eq.(\ref{ham-1}) that $\hat{c}_i\rightarrow -\hat{d}^\dagger_i$ if $i$ is odd, $\hat{c}_i\rightarrow \hat{d}^\dagger_i$ if $i$ is even and $\alpha \rightarrow \alpha+ 1/2$, then Eq.(\ref{ham-1}) keeps the same form if we exchange $t$ and $\Delta$, which means $t$ and $\Delta$ make same contribution to extension and localization of this system. The range of critical region shouldn't change if we exchange $t$ and $\Delta$, so we can determine the critical region $2||t|-|\Delta||\leq V \leq2||t|+|\Delta||$ just from our choice of $\Delta>0$ and $t>\Delta$, no matter how large $t$ and $\Delta$ are.

%\clearpage
%%%%%%%%%%%%%%%%%%%%%%%%%%%%%%%%%%%%%%%%%%%%%

\end{document}